\documentstyle[12pt,epsf]{article}

\newcommand{\eps}{\epsilon}
\newcommand{\ds}{\displaystyle}
\newcommand{\ra}{\rightarrow}
\newcommand{\be}{\begin{equation}}
\newcommand{\ee}{\end{equation}}
\newcommand{\bea}{\begin{eqnarray}}
\newcommand{\eea}{\end{eqnarray}}
\newcommand{\ci}{\cite}
\newcommand{\bi}{\bibitem}
\newcommand{\nono}{\nonumber \\}

\newcommand{\dd}{\partial}

\newcommand{\bfo}{\mbox{\boldmath$\omega$}}

\newcommand{\bfnabla}{\mbox{\boldmath$\nabla$}}
\newcommand{\half}{\frac{1}{2}}

\newcommand{\vv}{\vec{\bf{v}}}

\newcommand{\x}{{\bf x}}

\def\dal{\,\lower0.3ex\vbox{\hrule\hbox{\vrule\kern2pt\vbox{\kern4pt\kern4pt}
\kern2pt\vrule}\hrule}\,}

\begin{document}

\title{\sl A nonlinear differential approach to the Saffman-Taylor finger}
\vspace{1 true cm}
\author{G. K\"albermann, and R. Wallach\\Soil and Water dept., Faculty of
Agriculture, Rehovot 76100, Israel}
\date{}
\maketitle
\begin{abstract}

Nonlinear time-dependent differential equations 
for the Hele-Shaw, Saffman-Taylor problem are derived.
The equations are obtained using a separable ansatz expansion for the
 stream function of the displaced fluid obeying a Darcian flow.
Suitable boundary conditions on 
the stream function, provide a potential term for the nonlinear equation.
The limits for the finger widths derived from the potential and
 boundary conditions are $1>\lambda>\frac{1}{\sqrt{5}}$, 
in units of half the width of the Hele-Shaw cell, in accordance
with observation.
Stationary solutions with no free phenomenological parameters
are found numerically.
The dependence of asymptotic finger width on the physical
parameters of the cell compares satisfactorily with experiment.
The correct dispersion relation for the instabilities is obtained from the
 time dependent equation.

\end{abstract}

\newpage
\section{\label{intro} Introduction}

The Hele-Shaw cell experiment of a less viscous fluid
displacing by a more viscous one is the paradigm of fingering phenomena\ci{pel}.
The time-honored, work of Saffman and Taylor\ci{st}
showed both experimentally and theoretically that fingers with
beautifully regular profiles arise in the Hele-Shaw cell.

In the intervening years a large body of works has added to our knowledge
of the fingering phenomenon in various branches of the dynamics of continuous 
media, such as dendritic growth, directional solidification, diffusion-
initiated aggregation, flame front propagation, electromigration, as well
as fingering in porous media. The porous medium problem
 prompted the initial research of fingering phenomena.
The topic is more than ever now, of the utmost importance in problems of
transport in saturated and unsaturated soils, groundwater pollution, etc.
\footnote{The website http://www.maths.ox.ac.uk/~howison/Hele-Shaw
/helearticles.bib
, cited in ref.\ci{ho20} carries an extensive (more than 600 papers), 
up to date list of references on the Hele-Shaw problem.}
Moreover, research in the field is now evolving into more complicated
geometries, like that of a spherical shape cell.\ci{pari}

Despite considerable efforts, the fingering phenomenon remains 
in many aspects uncharted territory. As Tanveer\ci{tan} described it
very lucidly, the phenomenon defies intuition. 
In the words of Tanveer\ci{tan} , the disproportionally large influence of small
effects, like local inhomogeneities, thin film effects, and 
the crucial role of surface tension make the theoretical
description very difficult. So crucial is the surface tension in
the fingering phenomenon, that the problem is ill-posed
without it\ci{ho86}.
The original work of Saffman and Taylor\ci{st} observed 
fingers in a cell with a less viscous fluid (oil, air, water), penetrating into 
a more viscous fluid (oil, glycerine). An initial instability
develops into a finger or several competing fingers.
Saffman and Taylor\ci{st} compared the experimentally photographed finger
profiles to a
theoretical prediction. 
The formula was obtained by taking advantage of 
the analytical properties of the
complex fluid potential (potential and stream function).
Reasonable boundary conditions, and neglect of the surface tension, gave
good agreement with the experimental profiles only in the
limit of very large capillary number (or alternatively small surface tension).
A drawback of this theoretical approach lies in the absence of
surface tension that, causes mathematical ill-posedness
of the problem.
This fact is reflected in the absence of a limitation for the
asymptotic size of the finger. Contrarily to the measured profiles, that
were always found to be limited by approximately one half of the cell
width, the theoretical expression showed no such lower bound.

McLean and Saffman\ci{mc} addressed the problem including 
surface tension.
Similarly to Saffman and Taylor they emphasized the role of analyticity.
Tanveer\ci{tan} has shown that the approach of McLean and Saffman\ci{mc}, 
is equivalent to an expansion
in a parameter related to the finger width, capillary number and aspect ratio. 
The expansion fails at the tail of the finger and higher order 
terms are needed. The results of McLean and Saffman\ci{mc} nevertheless
yielded profiles that matched very well the front (nose) part of the finger.
Also they found a limit of $\lambda>\frac{1}{2}$ to the finger width.
Their curve for the dependence of the finger width on the capillary number 
show the right trend, albeit below the measured values.
There appear to be, however, other branches of solutions (two at least)\ci{van},
that compete with the branch found by McLean and Saffman\ci{mc}.
These branches are presumably unstable as compared to
the one found by McLean and Saffman\ci{mc}.

More accurate experiments with many different aspect ratios ( the
ratio of width of the channel with respect to its thickness) showed that
 the limit is more likely around
$\lambda >0.45$ of the cell width\ci{tab}.

Amongst the phenomenological approaches to the problem, the remarkable work of
Pitts\ci{pit}, took advantage of the observed dependence of the curvature
of the finger on the angle, to obtain an analytical scale covariant
expression for the finger shape that fitted measured values (by Saffman
and Taylor and by Pitts himself) extremely well, especially for finger
widths smaller than $\lambda\approx 0.8$. For wider fingers, 
the profile function
was found to miss the measured finger by a small amount.
To this day, there is no other stationary finger analytical 
expression in the literature that fits the data.
Pitts derived also an expression for the dependence of the
finger's asymptotic width on the capillary number that includes a 
phenomenological (fitted) parameter taking into account the
fluid films left behind by the passing finger. The
fit was also very successful. 
The theoretical basis for the phenomenological assumptions
used by Pitts are nevertheless unclear.

DeGregoria and Schwartz\ci{deg} used a boundary integral Cauchy technique to 
investigate the production and propagation of fingers in the
Hele-Shaw cell.
Tryggvason and Aref\ci{tri} used a {\sl vortex in line}
method to determine the relationship between 
finger width and flow parameters. Both works, as well as the stationary
calculation of McLean and Saffman\ci{mc} give similar results
concerning the finger width dependence on capillary number.
A marked improvement is found in the numerical calculations
of Reinelt\ci{rei}.
  
The intention of the present work, is 
to offer yet another approach to the fingering
phenomenon. We will derive a nonlinear differential equation for the
finger in the Hele-Shaw cell.
The equation
resembles the Korteweg-de Vries (KdV) soliton equation\ci{rem}.
It will be quite rewarding to find several features emerging from 
this method, like the ability to reproduce stationary finger profiles 
 accurately, and the 
possibility to deal with transient phenomena such as the development and
competition between fingers and the treatment of tip splitting.
In the present work we will focus on the justification of the 
nonlinear soliton-like equation, and its stationary solutions. 

\section{\label{streamf}Basic equations}

The basic equations governing the fluid flow are described in many works
\ci{pel,tan}.
We here reproduce them for the sake of completeness, and add our view
on the problem.
The Hele-Shaw creeping flow equation that neglects inertia effects and
integrates over the transverse dimension of the cell is\ci{st}

\be\label{creep}
\vv=-\frac{b^2}{12~\mu}\vec{\bfnabla} p
\ee

where $\vv$ is the viscous fluid velocity in the plane of the cell,
{\sl b} is the thickness of the cell, $\mu$ the viscosity, and {\sl p},
the pressure in the fluid.
Even at this level, there are hidden assumptions that are not clearly
valid in all situations. The integration over the third dimension indeed
helps in reducing the problem to a more manageable set of equations.
However, the true and real problem is inherently
three-dimensional\ci{tri}.
The existence of films of fluid left behind the finger, puts this
aspect in evidence.
The finger thickness is not constant comparing tail and nose. It is thinner
at the nose and thicker at the tail. We are therefore already 
reducing dramatically the complexity, and richness, of the phenomenon.
The price is paid at the time of comparing the measured finger widths as
a function of the parameters of the flow to the theoretical values.
As demonstrated by Reinelt\ci{rei}, a careful treatment of 
films of fluid by means of a three (or four) domain splitting of the Hele-Shaw
cell, changes dramatically the agreement between theory and experiment.

Originally eq.(\ref{creep}) was supposed to imitate Darcy's
law for the flow in porous media\ci{st}.
The phenomenon is nevertheless of a general nature.
It belongs to a large class of curved front propagation
 phenomena spanning such diverse topics
as dendritic solidification and flame combustion.

The displacing fluid also obeys a Darcian law as above, however, and due
to the lower viscosity, it is usually ignored and the pressure is
assumed constant inside the finger. This assumption is also not perfectly
justified, but, seems an economical working hypothesis, especially when the
displacing fluid is a gas.

The above equation has to be augmented by boundary conditions
at the interface. Here too the court is open to a variety of conditions,
mainly due to the ignorance of the third dimension.
Basically, the main boundary condition is the one of the pressure
difference between both sides of the interface.
Without corrections of thin films\ci{rei,park} the condition reads\ci{tan}

\be\label{cond1}
\Delta p=\frac{2~T}{b}~cos\gamma+\frac{T}{R}
\ee
 
where, {\sl T} is the surface tension parameter of the Young-Laplace formula,
{\sl R}, the radius of curvature in the plane of the cell and $\gamma$, the
contact angle between the finger and the cell wall in the transverse {\sl b}
direction.
Many corrections were proposed to the above equation. 
Park and Homsy\ci{park} and
Reinelt\ci{rei}, developed a systematic expansion of the effective surface 
tension parameter as a function of the capillary number {\sl Ca=}$\frac{\mu~V}
{T}$, where {\sl V} is the asymptotic velocity of the fluid far ahead of the 
finger that is conventionally related by flux conservation (and without 
considering Poisseuille flow, or else no-slip condition) to the velocity of
the finger {\sl U} by $V=U\lambda$. Here $\lambda$ is half the width of the
finger tail after it has fully developed and other competing fingers
dissolved.

The term in (\ref{cond1}) depending on {\sl b} is usually dropped, because the
flow depends on the gradient of the pressure and a constant term does not 
matter. However, this is not valid either. Two effects spoil the constancy,
 the variation of contact angle $\gamma$ and, the change of finger
thickness as one proceeds from nose to tail, or viceversa.
It is easy to see that the latter effect is very significant due to the
smallness of {\sl b} as compared to the width of the cell. A tiny change
in the thickness of the finger carries over to 
a large modification to the pressure.
However, the two-dimensional approach can not deal with this matter
consistently. 

The contribution of the
thinning out of the finger when coming from the tail to the nose adds up 
to the effect of the surface tension in the plane of the cell. One can then
hope that an effective surface tension might somehow lump
up the effect. 
However, this expectation seems to be too optimistic.

Pitts\ci{pit} found the need to introduce a phenomenological factor to
to account for the difference in pressure inside the fluid between tail
and nose that reflects the thinning out of the finger.
We will not delve here into the issue, but we do recognize that
without a reliable treatment of this matter, only the trend of
the curve of finger width versus capillary number is of interest.
In particular we will follow the same path that ignores the first term
in eq.(\ref{cond1}) and use instead an effective (unknown)
surface tension. 
For large surface tension,
i.e. fingers that cover almost the whole cell the treatment will 
be correct. The narrower the finger as compared to the
channel width, the stronger the effect of films of fluid left behind\ci{rei}
Therefore, without a three-dimensional treatment we have to be
satisfied with results for wide fingers and small capillary numbers.
Some comments on a possible parametrization that might account for the
loss of information from the third dimension will be provided later.

Equation (\ref{cond1}) reduces then to

\be\label{cond11}
\Delta p=\frac{\tilde T}{R}
\ee

with ${\tilde T}$ an effective surface tension parameter.

To complete the set of equations we assume the fluids to be immiscible.
This means that there is no interpenetration. The displaced fluid
does not cross the boundary. Its velocity is tangential to the boundary.
The boundary is then along a constant stream-line.
For a frame of reference with the finger at rest it becomes

\be\label{cond2}
\alpha(\vv)=tan^{-1}(\eta')
\ee

where $\alpha$ is the angle between the fluid velocity and the axis of 
propagation of
the finger (x axis) and $\eta'$ is the derivative to the finger (y axis)
with respect to the x axis.
The above condition is equivalent to the much simpler 
one of $\Psi=constant$ on the finger, where $\Psi$ is the
stream function for the viscous fluid\ci{st,mc}
, as well as to the condition of
continuity of the velocity in the normal direction to the interface
for a moving finger.
Note also that the treatment of Reinelt\ci{rei} being structured in 
three different regions, with the front inside an intermediate zone, does
allow for the crossing of fluid through the front. As it is our aim
to provide an alternative differential approach, we will not
deal with such complications at this stage.

To the above equations one has to add the continuity equation for
an incompressible fluid of constant density, $\ds \vec{\bfnabla}\cdot\vv=0$.

The transformation of the Hele-Shaw cell problem to a two-dimensional
situation permits the use of analytic functions techniques.
Various authors took advantage of this feature either to
expand the complex flow function in terms of the complex
coordinate $z=x+i~y$ or viceversa\ci{st} , or
use analyticity conditions to approximate the solution\ci{mc,deg}
The potential $\Phi$ is defined by $\vv=\vec{\bfnabla}\Phi$ and the stream
function by $\vv=\vec{\bfnabla}{\x}\vec{\Psi}$. 
A single component of this vector is needed for flows
averaged over the cell thickness, $\Psi_b$, where
{\sl b} denotes the direction transverse to the cell.
Moreover and in accordance with the Hele-Shaw approach, this function has
to be independent of the transverse coordinate. 
The Cauchy-Riemann equations for the potential and stream function,
 imply that the complex potential $\omega=\Phi+~i~\Psi_b$, 
is analytic in {\sl z}.

Instead of working with the full complex potential and derive restrictions
based on analyticity, we here treat the problem by means of the
stream function.
The stream function is a vector, whose curl determines
the fluid flow. The incompressibility of the flow is then guaranteed.
Although the velocity potential $\Phi$ is a harmonic function of
the coordinates $x,y$, as can be readily obtained by applying
the incompressibility condition to Darcy's law of eq.(\ref{creep}), the
stream function $\vec{\Psi}$ does not obey the equation $\nabla^2\vec{\Psi}=0$
automatically. 
The stream function approach and the potential one differ. 
The use of the velocity potential $\ds \vv=\vec{\bfnabla}\Phi$
assumes a vanishing vorticity, as the curl of a gradient vanishes. 
It is our claim, that this assumption is wrong. 
The stream function approach is superior, as it does not suppose
 such a neglect of vorticity. The velocity potential $\Phi$ misses
 a crucial part of the problem. Therefore, it should not be considered as 
a trustworthy tool to find the finger equation of motion.
In this respect the use of the full potential 
$\ds \omega=\Phi+~i~\Psi_b$, is also not valid. 
Hence, the analyticity conditions
 derived from it implying a harmonic equation for $\Psi$ also, do not
 reflect the full reality of the problem. 

For example, if we consider $\Psi_b$,  that depends on x and y only, we have  
$\ds \vec{\bfnabla}\cdot\vec{\Psi}=0$, because $\Psi_b$ is the {\sl z}
component of the stream vector and it does not depend on {\sl z}. $\ds \Psi_b$
satisfies the incompressibility condition, by definition of the velocities.
Using $\ds \vec{\bfnabla}{\bf x}\vec{\bfnabla}{\bf x}\vec{\Psi}
=\vec{\bfnabla}\vec{\bfnabla}\cdot\vec{\Psi}-{\nabla}^2\vec{\Psi}$,
 it is clear that, $\Psi_b$ will obey the harmonic equation if the
 vorticity, 
$\ds \vec{\bfo}=\half\vec{\bfnabla}{\x}\vec{\bfnabla}{\x}\vec{\Psi}$, 
vanishes. 
The vorticity may be ignored, except at the interface\ci{tri}. However,
it is there that we need the most the value of the stream function.
Therefore we will not use the harmonic equation in the present work.

The boundary conditions alone will suffice
 for the determination of the stream function on the front, and, in turn, 
allow us to find the evolution equations for the finger.
However, we will not be able to find the stream function explicitly everywhere
in the fluid.

\section{\label{str1} Stream function approach}

The stream function will be determined by resorting to a perturbation 
expansion. The
expanded stream function will be made to obey the equations of
the flow as well as the boundary conditions.
The outcome will be a stream function that is known solely at the boundary.
This method will enable us to deduce the nonlinear differential equations.

We first renormalize the axes to 
$\ds x\ra{\tilde x}=
\frac{x}{w/2},~y\ra{\tilde y}=\frac{y}{w/2}$. As a consequence of
this rescaling, the finger
transverse dimension becomes less than one, 
and serves as an expansion parameter.

Two boundary conditions constrain the stream field.
The first one is the impenetrability of the cell wall and, the second
the immiscibility of the fluids. The latter amounts to
considering the interface between the fluids as a free surface.
Dropping the suffix {\sl b} and the tilde, we find

\bea\label{constraint1}
\frac{\dd\Psi}{\dd x}=0~,at~y=\pm1\nono
\eea
  
and 

\bea\label{constraint2}
\frac{\dd\eta}{\dd t}+\frac{\dd\Psi}{\dd y}\frac{\dd\eta}{\dd x}=
-\frac{\dd\Psi}{\dd x},~at~y=\eta
\eea

where $\eta(x,t)$ denotes the interface curve.
Eq.(\ref{constraint2}) may be easily derived by writing the 
differential of $\eta$ as $\ds 
d\eta=\frac{\dd\eta}{\dd t}~dt+\frac{\dd\eta}{\dd x}
~dx$ and, using the definitions 

\bea\label{vels}
v_x&=&\frac{dx}{dt}~=~\frac{\dd\Psi}{\dd y}\nono
v_y&=&\frac{dy}{dt}~=~-\frac{\dd\Psi}{\dd x}
\eea

For the stationary finger, in its rest frame we have 
$\ds\frac{\dd\eta}{\dd t}~=~0$. 

With this substitution, eq.(\ref{constraint2}) becomes

\bea\label{const5}
\frac{\dd\Psi}{\dd y}\frac{\dd\eta}{\dd x}=-\frac{\dd\Psi}{\dd x},~at~y~=~\eta
\eea

The differential of $\Psi$ 
$\ds d\Psi=\frac{\dd\Psi}{\dd y}~dy+\frac{\dd\Psi}{\dd x}~dx$, 
together with eq.(\ref{const5}) determine
 the unique solution for the stream function in the stationary
case to be 
\be\label{psi0}
\Psi_0~=~constant
\ee
Without loss of generality, this constant may be taken to be equal
to zero, because the stream function enters the calculations only
through its derivatives.

For a finger in motion along the {\sl x}
axis, even with a time dependent
velocity, but still in the creeping flow approximation that neglects
kinetic terms, we can write the solution to be formally 

\be\label{const1}
\Psi(x,y,t)=\Psi_0(x,y)-\int{dx~~\frac{\dd\eta}{\dd t}}
\ee

Eq.(\ref{const1}) solves eq.(\ref{constraint2}), with $\Psi_0$ given by 
the appropriate solution of eq.(\ref{const5}).

At the front, and only there, we can rewrite the equation as

\be\label{const12}
\Psi(t)=\Psi_0-\int{dy~\frac{\dd\eta}{\dd t}/\frac{\dd\eta}{\dd x}}
\ee

Where $\eta(x,t)$ is understood as a function of {\sl x and t} only. In the
general case, $\Psi$ is no longer a constant on the curved front.

\vspace{ 1 cm}
Consider the stream function in the finger rest frame. The finger
is traveling in the positive {\sl x} direction at constant speed {\sl U}.
Moreover, at long distances ahead of the finger the fluid is assumed to flow
with a constant velocity {\sl V}. Usually, the no-slip condition cannot be
imposed in the Hele-Shaw cell and consequently the velocity {\sl V} is 
considered to be the same all over the cell including at the lateral boundary.
This could be a good assumption provided the viscosity of the fluid is small,
or else the lateral edge is not considered as the boundary. 
However, for large capillary numbers, or large viscosities, this
procedure seems unreasonable. Boundary layers next to surfaces are
patent in common phenomena even for moderately viscous fluids.
We opt for a more conservative approach and demand the
stream line next to the lateral edge to carry a velocity {\sl V} that
is {\sl not} the asymptotic velocity of the fluid ahead of the finger.
It is a parameter that depends on the flow properties. 
For capillary number tending to infinity it has to be equal to zero.
Large capillary numbers may be implemented by
using very viscous fluids. For such fluids, the no-slip condition is a must.
Therefore in the limit of infinite capillary number {\sl V=0}.
In the opposite limit, i.e. no viscosity at all (finger width going to
the full cell width) it is equal to the finger velocity. Hence
we can assert $0<V<U$. Reinelt's equations\ci{rei} carry
an extra term at the lateral edge that allows the velocity
{\sl V} to be variable.

To our knowledge there have not been any
experimental investigations of the fluid flow as a function
of distance from the cell edge. In light of the previous
discussion, this measurement could be of relevance.

Expanding the only component of the 
stream function, with a separable ansatz, antisymmetric in {\sl y}, so that
the finger is symmetric, 
and in the finger rest frame, we find 
(dropping from now on the tilde on {\sl y}, i.e. we work with rescaled
coordinates)

\bea\label{stream}
\Psi(x,y,t)=y~(V(x,t)-U(x,t)+A(x,t)+B(x,t)~y^2+C(x,t)~y^4+...)
\eea

with {\sl A, B, C} unknown functions. We will find below
that the expansion will effectively be in terms of both {\sl y}
and $1-y^2$. It is then expected to be a reasonable
 approximation in all the cell gap, except perhaps around
$y\approx 0.5 $

Eq.(\ref{constraint1}), as well as the constancy of the fluid velocities, for
the stationary finger in its rest frame imply

\bea\label{dem}
v_y&=&-\frac{\dd\Psi}{\dd x}=0~,at~~y~=~\pm 1\nono
v_x&=&\frac{\dd\Psi}{\dd y}=~V-U~,at~~y~=~\pm 1
\eea

Inserting eq.(\ref{dem}) in eq.(\ref{stream}), we are able
to fix the functions {\sl B and C} in terms of {\sl A}.
Explicitly
\bea\label{sldem}
0&=&A+3~B~+5~C~\nono
0&=&~A'+B'+C'
\eea

primes denoting derivatives with respect to {\sl x}.

Recalling that constants are irrelevant for the stream function, 
the unique solution to this order becomes 

\bea\label{soldem}
B&=&-2~A\nono
C&=&A
\eea

The stream function of eq.(\ref{stream}) with the conditions of 
eqs.(\ref{sldem},\ref{soldem}) reads

\be\label{stream1}
\Psi(x,y)=(V-U)~y+A(x)~y~(1-y^2)^2+...
\ee

Eq.(\ref{stream1}) demonstrates that the expansion is a dual one
 in $y$ and, $1-y^2$, as stated above.
In eq.(\ref{stream1}),
we have neglected higher order terms and also ignored possible
contributions to the stream function from the dependence of the
velocities {\sl V, U} on distance. This is clearly not a good
approximation for large capillary number as well as for the nonstationary
problem. 

If we expand the stream function to the next order
\bea\label{streamh}
\Psi(x,y,t)=y~(V(x,t)-U(x,t)+A(x,t)+B(x,t)~y^2+C(x,t)~y^4+D(x,t)~y^6+...)
\eea

we need an extra condition to determine the function {\sl D}.
One possible constraint to fix this function, may arise from a model of 
the vortex structure on the curved front. 
The simplest one - although not the most realistic one-, 
consists in assuming a constant vorticity on the interface,
with the vorticity defined by 
$\ds \vec{\bfo}=\half\vec{\bfnabla}{\x}\vec{\bfnabla}{\x}\vec{\Psi}$.
This procedure introduces a new parameter, the vorticity value $ \omega$.

In order to stay as low as possible in the number of free parameters,
we prefer to 
remain within the expansion to $O(y^6)$ with $y <1$. To this order
the two conditions of eq.(\ref{constraint1},\ref{constraint2}), are sufficient.
The constraints we take advantage of, 
depend on the stream function and its first derivative.
Higher order expansions as in eq.(\ref{streamh}) and further constraints,
 as the vorticity, amount to an expansion in second and higher order derivatives
of the stream function. Due to the fact that,
the even polynomials in {\sl y} span a complete
set in the rescaled variable $0< y<1$ for the stationary finger case,
this expansion presumably converges.
The sixth order term is expected to be smaller
than the fourth order one by factors of the form $\lambda^2$, with
$\lambda<1$.
Although the exact dependence in $\lambda^2$ is not known at this stage, it
appears a sound working hypothesis to truncate the expansion to the
sixth order and work with the expression of eq.(\ref{stream1}).
The quality of the finger shapes obtained bellow gives us more confidence 
in this hypothesis.

\section{\label{equ}Stationary and time-dependent 
nonlinear differential equations for the finger}

On the curved front, in the finger rest frame, 
with the figer nose located at the center of coordinates, the condition
$\Psi~=~constant$ of eq.(\ref{psi0}) becomes 

\bea\label{cond4}
\Psi=0,~~at~y=\eta
\eea

Inserting eq.(\ref{cond4}) in
 eq.(\ref{stream1}), $A(x)$ is determined at the front to be

\bea\label{ax}
A(x)=\frac{U-V}{{(1-\eta^2)}^2},~ at~~y=\eta(x)
\eea

Using eqs.(\ref{vels},\ref{stream1},\ref{ax}),  
the velocity of the fluid at the static finger, 
the only place where we can determine {\sl A} in closed form, becomes

\bea\label{vel}
v_{x,static}&=&(V-U)+A(x)~(1-\eta^2)~(1-5~\eta^2)\nono
v_{y,static}&=&-\frac{\dd A(x)}{\dd x}~\eta~{(1-\eta^2)}^2\nono
A(x)&=&\frac{U-V}{{(1-\eta^2)}^2}
\eea

with $\eta$, a function of x. The fluid still flows around the finger
in its rest frame, hence the name {\sl static}.

The rescaled curvature in eq.(\ref{cond11}) is 

\bea\label{curv}
\frac{1}{R}=\frac{\eta''}{(1+\eta'^2)^{\frac{3}{2}}}
\eea

Where primes denote derivatives with respect to x. 

The sign of the curvature is the appropriate one.
This can be seen by using eq.(\ref{creep}). The left hand side of
the equation is a the positive velocity of the finger, chosen
 here to move from left to right, therefore
the curvature has to decrease along {\sl x}. In the upper part
of the finger the curvature is negative and becomes more so
as we proceed along {\sl x}. Therefore the right hand side
 of eq.(\ref{creep}) is positive with the positive sign in eq.(\ref{curv}).

Eq.(\ref{curv}) is the standard expression 
for the curvature, and, it is easily derivable from the definition
of the arclength and the corresponding angle for differential
increments. 

Equations (\ref{creep},\ref{cond11},\ref{curv}) after rescaling imply

\bea\label{diff0}
v_x=-~\frac{\tilde T~b^2}{3~w^2\mu}\frac{\dd}{\dd x}
\bigg[\frac{\eta''}{(1+\eta'^2)^{\frac{3}{2}}}\bigg]
\eea

The stationary finger equation is obtained now by
evaluating the derivative in eq.(\ref{diff0}), with 

\be\label{vx}
v_{x,static}=(V-U)~\frac{4~\eta^2}{1-\eta^2}
\ee

from eq.(\ref{vel}); and transforming back to the rest frame
of the cell, with the finger in motion, $\ds v_x~=~v_{x,static}+U$.

After some straightforward algebraic manipulations we find

\bea\label{diff}
0&=&\eta_{xxx}-3~\frac{\eta_{xx}
^2~\eta_x}{1+\eta_x^2}~+~(1+\eta_x^2)^{\frac{3}{2}}~\frac{W(\eta)}{4~B}\nono
W(\eta)&=&\frac{4~\eps~\eta^2+1-5~\eta^2}{1-\eta^2}
\eea
where the suffix indicates differentiation with respect to {\sl x}.

In terms of the arclength measured from the tail of
the finger, the equation becomes

\bea\label{diffs}
0&=&\eta_{sss}+\frac{\eta_{ss}
^2~\eta_s}{1-\eta_s^2}+(1-\eta_s^2)~\frac{W(\eta)}{4~B}\nono
\eea

In eqs.(\ref{diff},\ref{diffs}), $\ds \eps=\frac{V}{U}$.
As was found in previous works\ci{mc,tab}
the only parameter entering the equation is $\ds \frac{1}{B}
=\frac{12\mu~U~w^2}{{\tilde T}~b^2}$
with $w$ being the width of the Hele-Shaw cell and all the variables are scaled
to units of half this length.
The existence of a single parameter is clearly insufficient
as demonstrated by Reinelt\ci{rei}.
Customarily $\eps$ of eqs.(\ref{diff},\ref{diffs}) is equated to 
$\lambda$, half the finger width. We here take it as a parameter as explained
in section (\ref{str1}). Asymptotically far back at the tail
we must have $W(\eta)~=~0$, or, 
$\ds \epsilon=\frac{5~\lambda^2-1}{4~\lambda^2}$.
The parameter $\epsilon$, is then determined by the solutions to
the equations and is not a phenomenological parameter.

The potential {\sl W} guides the propagation of the finger. The equation is
of third order in the spatial derivatives as the Korteweg-deVries
equation\ci{rem}. Third order differential equations for
interface propagation are well known in the literature. Some notorious
examples are: Landau and Levich\ci{land}, Bretherton\ci{bre}, 
and Park and Homsy\ci{park}.

The third order equation (\ref{diff}) can be transformed to a more manageable
second order one in terms of the angle tangent to the curve, 
with $\eta$ obtained by integration. For $ds$ starting at the tail, 
where $\ds\theta\approx\pi$, $\ds ds~cos(\theta)=~-dx$.

Equation (\ref{diffs}) now reads

\bea\label{diff2}
0&=&\frac{{\dd}^2\theta}{\dd s^2}-cos(\theta)~\frac{W(\eta)}{4~B}\nono
\eta&=&\lambda-\int{ds~sin(\theta)}
\eea

With $W(\eta)$ defined in eq.(\ref{diff}).

\vspace{2 cm}

The time dependent equation is obtained by using eq.(\ref{const12}), 
and recalling the definition of $v_x$ in terms of the stream function of
eq.(\ref{vels}). For the velocity in the {\sl x} direction, we now have

\bea\label{dyna}
v_x=-\frac{\eta_t}{\eta_x}~+~U~+~v_{x,static}
\eea

Rescaling as before the length coordinates $\eta$, and {\sl x} 
by $\ds \frac{w}{2}$,
and the time by $\ds t\ra\frac{t~U}{w/2}$, and inserting eq.(\ref{dyna}) in 
eq.(\ref{diff0}), we obtain the time dependent differential equation

\bea\label{time}
0&=&\eta_{xxx}-3~\frac{\eta_{xx}
^2~\eta_x}{1+\eta_x^2}~+~(1+\eta_x^2)^{\frac{3}{2}}~
\frac{1}{4~B}\bigg[{W(\eta)-\frac{\eta_t}{\eta_x}}\bigg]
\eea

In terms of the arclength eq.(\ref{time}) becomes

\bea\label{times}
0&=&\eta_{sss}+\frac{\eta_{ss}
^2~\eta_s}{1-\eta_s^2}+(1-\eta_s^2)~\frac{1}{4~B}\bigg[
W(\eta)-\sqrt{1-\eta_s^2}\frac{\eta_t}{\eta_s}\bigg]
\eea

While for the angle tangent to the front, the 
time-dependent equation (\ref{time}) reads

\bea\label{time2}
0&=&\frac{{\dd}^2\theta}{\dd s^2}-cos(\theta)~\frac{1}{4~B}
\bigg[W(\eta)+cos(\theta)~{\frac{\dd\theta}{\dd t}}/
{\frac{\dd\theta}{\dd s}}\bigg]\nono
\eta&=&\lambda-\int{ds~sin(\theta)}
\eea

To gain confidence in the equations, we proceed to show that
the time dependent equation (\ref{time}) gives the correct
dispersion relation for the perturbation of a planar front.
We will find that the instability parameter in time, here
called $\sigma$, will obey a dispersion relation of the form
$\ds \sigma=|\alpha|~(1-\lambda\alpha^2)$, with $\alpha$ related to
the wavenumber of the perturbation and $\lambda$ a parameter
 proportional to the surface tension. The velocity, or pressure
gradient, destabilizes and the surface tension stabilizes.
We will connect to Chuoke et al.\ci{chuoke} in the derivation.

We will consider a perturbation protruding from a moving
front with velocity U. Eq.(\ref{time}) is firstly rewritten around 
$y\approx 0$

\bea\label{timep}
0&=&\eta_{xxx}-3~\frac{\eta_{xx}
^2~\eta_x}{1+\eta_x^2}~+~(1+\eta_x^2)^{\frac{3}{2}}~
\frac{1}{4~B}\bigg(1-\frac{\eta_t}{\eta_x}\bigg)
\eea

This equation is equivalent to

\bea\label{timepp}
\frac{\dd}{\dd x}~\chi&=&0\nono
\chi&=&4~B\frac{\eta_{xx}}
{(1+\eta_x^2)^{\frac{3}{2}}}~+x~-\int{dx~\frac{\eta_t}{\eta_x}}
\eea

Only around $y\approx 0$, the equation can be integrated once analytically.

Now we need an ansatz for the solution. Following Chuoke et al.\ci{chuoke}
 we use the expression

\bea\label{ansatz}
x(y,t)&=&\beta(\alpha)~e^{\phi}\nono
\phi&=&\alpha~y+\sigma~t
\eea

$\beta$ is the amplitude of the perturbation, that is taken as dependent on 
$\alpha$ due to the nonlinearity of the equation, as Chuoke
et al.\ci{chuoke} assumed. (See below their equation [4]).

The integral in eq.(\ref{timepp}) may be rewritten as

\bea\label{int1}
I=\int{dx \frac{\eta_t}{\eta_x}}~=~\int{x_y~x_t~dy}~~~at~~~y=\eta
\eea
Evaluating the integral (\ref{int1}) with eq.(\ref{ansatz}) we find

\bea\label{int2}
I&=&\frac{\sigma}{2}~\beta^2~e^{2\sigma~t}~(e^{2\alpha y}-1)\nono
&\approx&\sigma~\beta^2\alpha~y
\eea

The curvature may be calculated now by using 

\bea\label{curv2}
\frac{\eta_{xx}}{(1+\eta_x^2)^{\frac{3}{2}}}
&=&-\frac{x_{yy}}{(1+x_y^2)^{\frac{3}{2}}}\nono
&\approx&-x_{yy}\nono
&=&-\alpha^2~x
\eea

Using eq.(\ref{curv2}), the integral of eq.(\ref{int2}), and,
 {\sl x} of eq.(\ref{ansatz}), at $t=0$ to $O(y^2)$, 
as we are assuming a perturbative expansion, $\chi$ of eq.(\ref{timepp}),
 becomes

\bea\label{chi0}
\chi_0\approx(-4~B~\alpha^2\beta~+~\beta)~(1+~~\alpha~y)-~\sigma~\beta^2
\alpha~y
\eea

The constant piece of $\chi_0$ in eq.(\ref{chi0}), $-4~B~
\alpha^2\beta~+~\beta$, is irrelevant. This is the reason we do not need
 to specify the lower bound of the integral in eq.(\ref{int1}). We took it
as $y~=~0$. Any other choice will merely change the unimportant constant in 
 $\ds \chi$.

To satisfy eq.(\ref{timepp}), to the lowest order in $y$, eq.(\ref{chi0}) 
has to obey the algebraic condition

\bea\label{chiob}
-4~B~\alpha^3\beta~+~\beta\alpha-\sigma\beta^2\alpha=0
\eea

The third term in eq.(\ref{chiob}) is quadratic in $\beta$. 
Independence from the initial perturbation amplitude, 
requires $\beta$ to be a function of $\alpha$, as assumed in eq.(\ref{ansatz}).
 We can now determine this dependence to be
$\ds \beta\alpha~=~k$, with {\sl k}, 
a constant. $\beta$ is a positive number, therefore,
{\sl k} has to be positive for positive $\alpha$, or
 negative for negative $\alpha$.

Equivalently, the proportionality above can be written as

\bea\label{beta}
\beta=\frac{|k|}{|\alpha|}.
\eea

Inserting eq.(\ref{beta}) into eq.(\ref{chiob}) we find

\bea\label{chiobp}
-4~B~\alpha^3\beta~+~\beta\alpha-\sigma\beta\alpha
\frac{|k|}{|\alpha|}=0
\eea

Finally, redefining $\ds |\alpha|\ra\frac{|\alpha|}{|k|},~
\lambda=4~B~k^2$ we find the expected dispersion relation

\bea\label{sigma}
\sigma=|\alpha|~(1-\lambda~\alpha^2)
\eea

\vspace{2 cm}

The time-dependent equation was developed on the basis of
an antisymmetric stream function, eq.(\ref{stream1}), the solutions, may, 
nevertheless, lack any predetermined symmetry. 
This phenomenon may be related to the physical concept
of hidden symmetries and spontaneous symmetry breaking\ci{itz}.
Although the equations are
symmetric under a certain transformation, the solutions can break the
symmetry.
Boundary conditions and initial conditions are key factors
in determining the symmetries of the solutions.
Such is the case of an initial 
asymmetric perturbation to the front.
On the experimental side, a symmetric finger develops
only asymptotically at long times. The transient behavior 
is patently asymmetric.

\section{\label{solu}Stationary finger solutions}

Let us analyze the predicted finger half-widths $\lambda$, as a function
of the parameter B. As discussed in section (\ref{equ}), we expect the
parameter $\eps$ to be equal to one in the limit of $B\ra\infty$, 
that is possible to implement with a displaced fluid of zero viscosity,
but a density that still assures the instability and creation of a finger. In 
this limit the potential determines the finger half-width ( location at which
the derivatives of $\eta$ vanish) to be $\lambda=1$, as expected.
On the other end of $B\ra~0$, or infinite viscosity limit, we argued
that $\eps=0$ in the spirit of the no-slip condition for viscous fluids.
In this limit $\lambda=\frac{1}{\sqrt{5}}\approx 0.447$. 
Looking at the experimental work of Tabeling et al.\ci{tab}, especially 
figure 8, it is seen that for $\ds \frac{1}{B}\ra\infty$ 
the finger width for various aspect ratios tends 
to approximately $\lambda=0.45$.

Therefore we can be quite confident on the
 potential {\sl W}.
On the other hand, the dynamics as described
by the third order differential equation linear in the parameter {\sl B}
is certainly only approximately correct. 
In the same manner as the KdV equation is only
an approximate equation derived by expanding to
a certain order in the derivatives.
The need of higher order terms in the curvature is apparent from
the work of Brower et al.\ci{brow}.
As stressed in the introduction, 
a full three dimensional account of the fluid flow
is necessary in order to take care of the thinning out of the
finger in the transverse direction. Alternatively we could view
 the equation as essentially correct, with
 the dependence on {\sl B} nonlinear and unknown.
Consequently, we can trust the equation in the limit of small $\ds \frac{1}{B}$.

The numerical solution was started from positive $y=\lambda$ and followed 
around the finger to $y=-\lambda$. We employed
a fourth order Runge-Kutta algorithm.
We opted for the mixed
integrodifferential approach of eq.(\ref{diff2}), 
that proved easier to handle numerically.
The equation can be solved without this change of variables also with the
full third order equation (\ref{diff}).
 
For fixed {\sl B} we varied manually the value of the asymptotic half-width
(thereby fixing $\epsilon$), the slope $\theta'$ and the value
of $\theta$ less than $\theta=180^0$.
The equation is integrated both backwards and forwards and the parameters
are fine tuned until the shape of the finger looks as symmetric as possible
 $x(-\lambda)\approx~x(\lambda)$, 
and the nose tip shows up at exactly $90^0$. We inforced the symmetry condition
to an accuracy of around 1$\%$.
Moreover, only solutions where the angle decreases smoothly from the
tail forwards were accepted. We rejected meandering (even slightly so)
solutions that could appear as a many prongued finger.

This method of tedious search proved to be reliable.
Other algorithms, such as a shooting method, etc., ended
up in unreasonable, asymmetric solutions.

The literature does not abound in theoretically calculated
profiles of fingers.
We could hardly find any such solutions besides those of McLean and 
Saffman\ci{mc}, the phenomenological solutions of Pitts\ci{pit}, and,
 the original ones of Saffman and Taylor\ci{st}. The latter
obtained without the inclusion of surface tension effects.
It appears then, that it is not an easy task to predict properly
finger shapes.

\begin{figure}
\epsffile{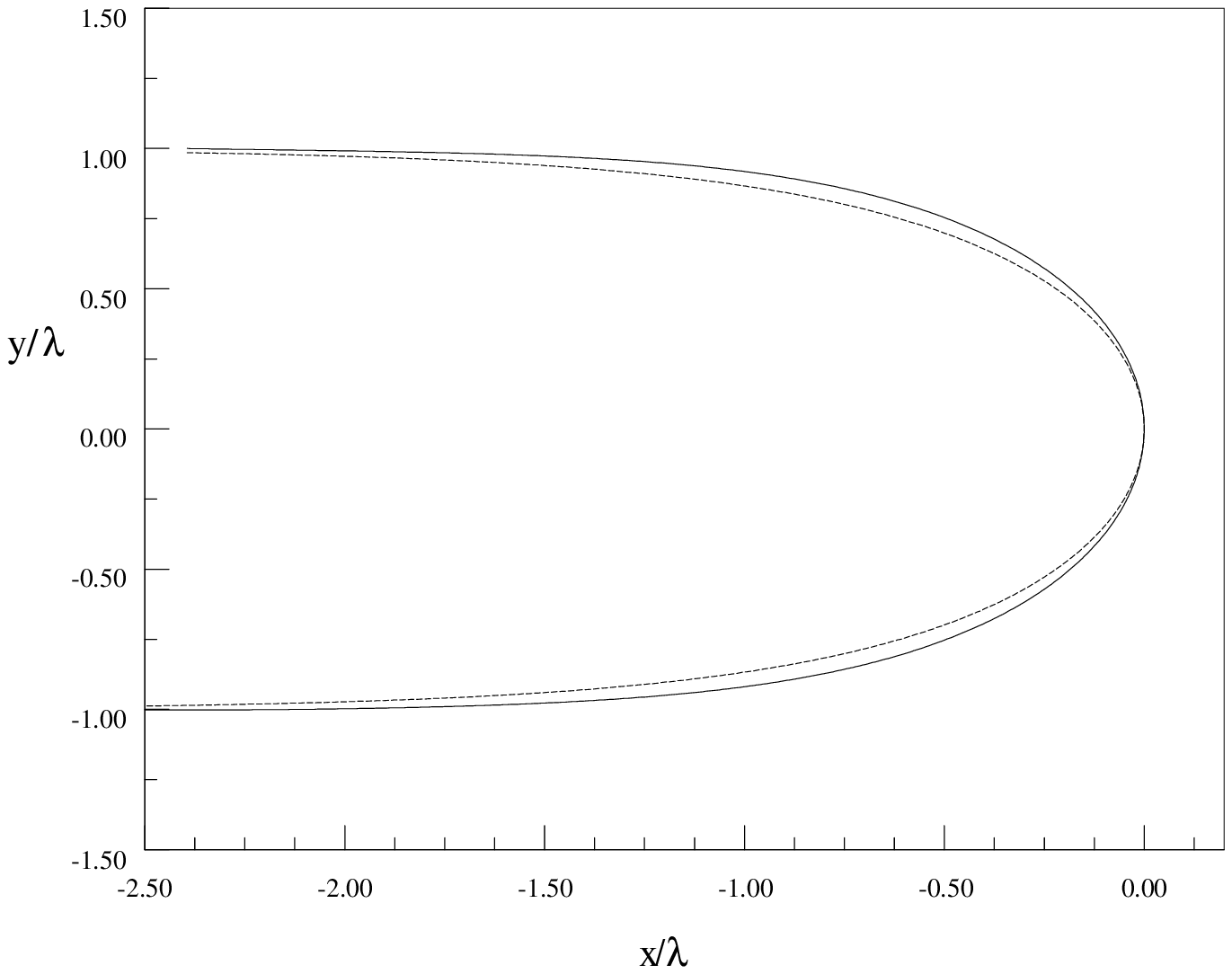}
\vsize=5 cm
\caption{\sl Rescaled finger profile for $\lambda=0.93$ at $\frac{1}{B}=12$. 
Numerical solution, solid line,
analytical solution of Pitts [10], dashed line.}
\label{fig1}
\end{figure}
We proceed to show 3 profiles found for finger half-widths 
$\lambda=0.93,0.82,0.66$. They are compared with the analytical solution 
of Pitts\ci{pit} that fits the data extremely well for finger half-widths
smaller than around $\lambda\approx 0.8$ and lies slightly 
below the data points for higher
values of $\lambda$. The figures are depicted for
rescaled coordinates in terms of $\lambda$ as has became customary\ci{mc,pit}.
\begin{figure}
\epsffile{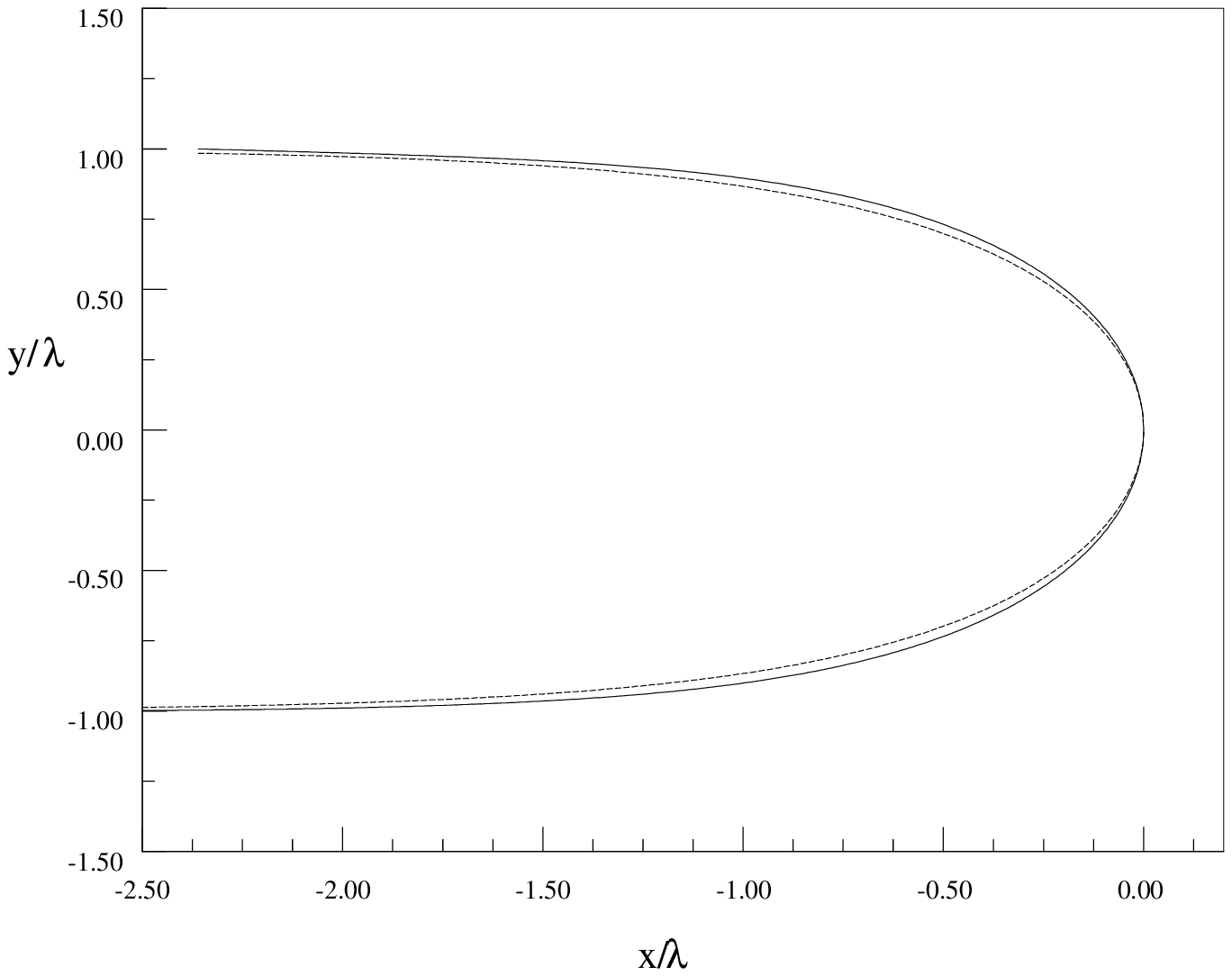}
\vsize=5 cm
\caption{\sl Rescaled finger profile for $\lambda=0.82$ at $\frac{1}{B}=20$. 
Numerical solution, solid line,
analytical solution of Pitts [10], dashed line.}
\label{fig2}
\end{figure}

From figures 1-3 it is clear that the solutions match the 
phenomenological solutions of Pitts $\ds cos(\frac{\pi}{2}\frac{~y}{\lambda})
e^{-\frac{\pi}{2}\frac{~x}{\lambda}}=1$, and are even superior to it for large
values of $\lambda$.
The dependence of the asymptotic finger half-width $\lambda$ 
on {\sl B}, is shown in figure 4.
The present results improve upon the calculations of
McLean and Saffman\ci{mc} in the range of {\sl B} for which we were able
to find reliable solutions, but still lie below the measured values
that were obtained from averaging the results of Tabeling et al.\ci{tab}.
for various aspect ratios.
Note however, that the results do not include finite gap corrections that
lift the curve upwards, whereas the results of McLean and Saffman\ci{mc}.
shown in the figure do adjust for such corrections. Without
these corrections the latter results fall much below the depicted curve.
As stressed above, finite film effects are extremely important,
as evidenced from the calculations made Reinelt\ci{rei}.

\begin{figure}
\epsffile{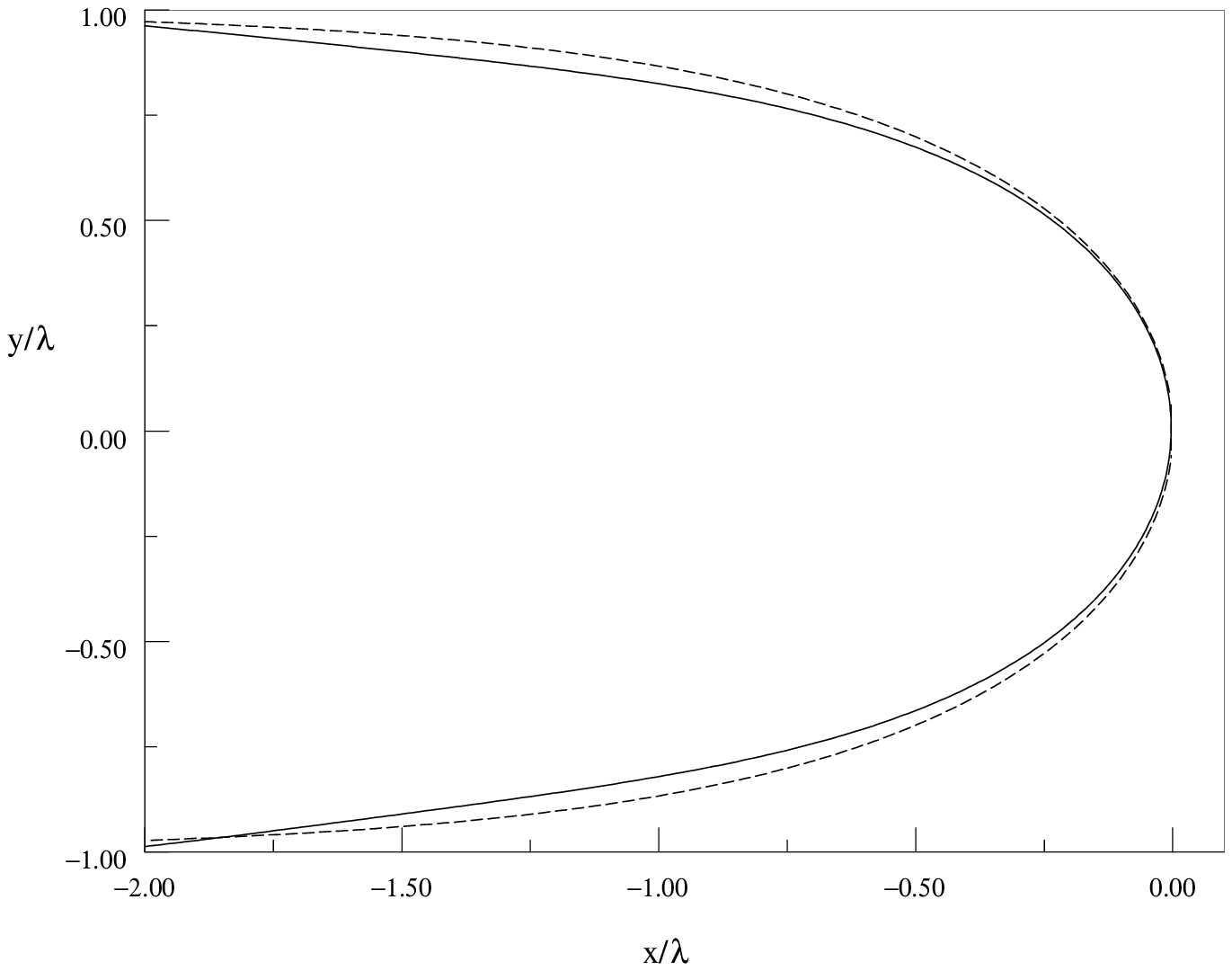}
\vsize=5 cm
\caption{\sl Rescaled finger profile for $\lambda=0.66$ at $\frac{1}{B}=48$. 
Numerical solution, solid line,
analytical solution of Pitts [10], dashed line.}
\label{fig3}
\end{figure}

\begin{figure}
\epsffile{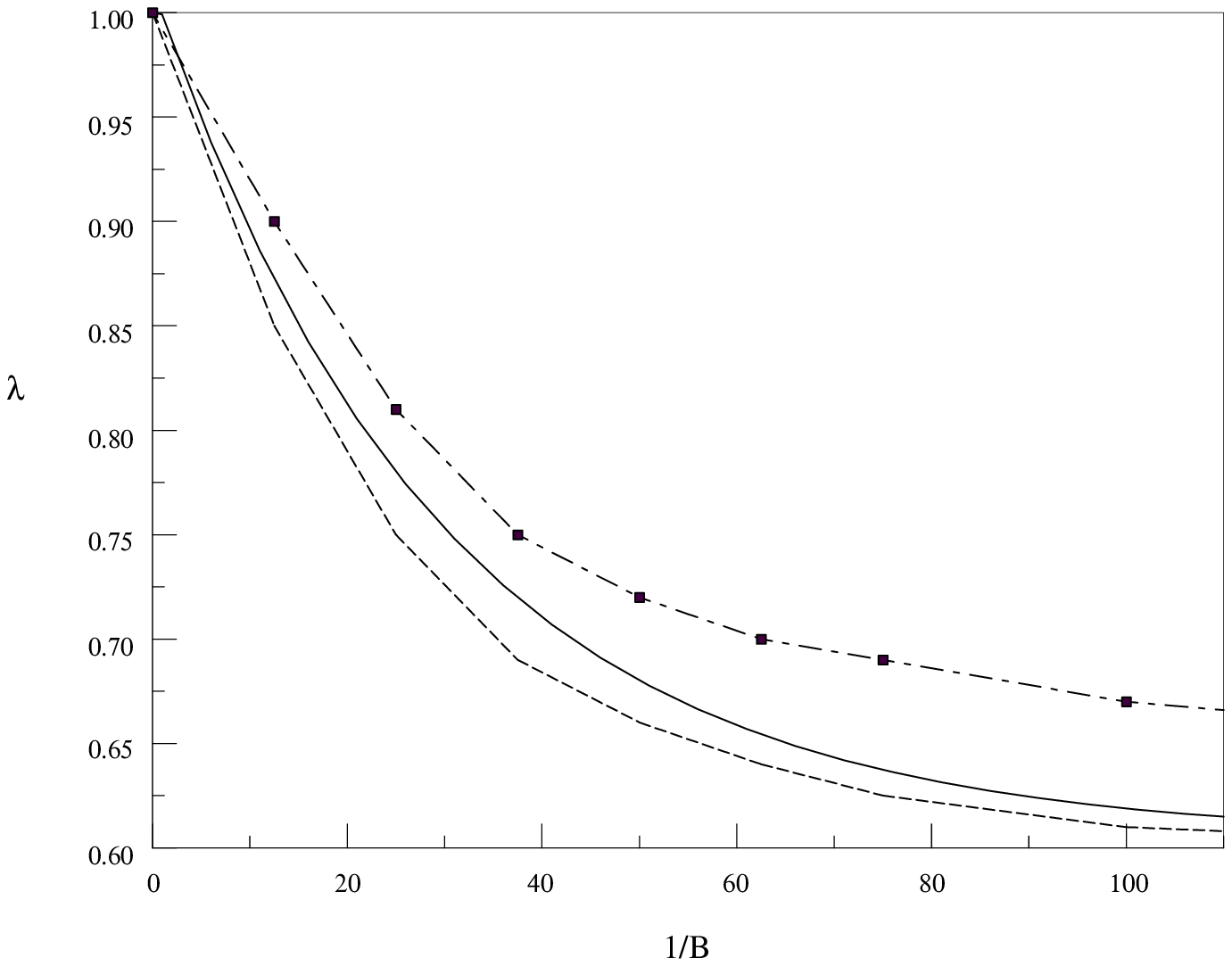}
\vsize=5 cm
\caption{\sl Asymptotic half-width of the finger, $\lambda$ as
a function of  $\frac{1}{B}$. Present work, full line,
theoretical results of McLean and Saffman [7] corrected for
finite film effects, dashed line, and,
 experimental results of Tabeling et al. [9], dash-dot line with squares.}
\label{fig4}
\end{figure}

We found numerically stationary finger profiles for {\sl B} $<$ 110.
Beyond this value, the large contributions introduced by the potential in 
eq.(\ref{diff2}), render the integration from the tail unstable.
This is not unexpected in light of the arguments provided in section 
(\ref{streamf}).
The curve of $\lambda$ versus $\ds \frac{1}{B}$ 
for large values of this parameter, can be fitted
by a function of the form $\ds B~e^{k~B}$, with {\sl k} a 
parameter. In future investigations 
higher order terms in the potential will be added, in the search of such a 
functional dependence in a self-consistent stream function approach.

Several attempts were made in the present study, 
to improve the algorithm by matching solutions
run from the tail and from the nose. However, starting from the nose
is very problematic, because of the singularity in the slope, that no
matter how, creeps up in any scheme, even after transforming
to better behaved variables.

The equation is extremely sensitive to the slope and initial angle
that must be close to $180^0$. Fortunately, it was found that for a 
a fixed value of {\sl B}, there corresponds a single value of $\lambda$; a 
unique finger solution.
The selection of the finger half-width is univocous, with all the numerical
methods employed.
However, we cannot rule out completely the existence of
other unstable branches not accessed by the numerical method, 
as found by Vanden-Broek\ci{van} for the McLean and Saffman\ci{mc} set of
integrodifferential equations.

In the course of the numerical investigation we found solutions
that correspond to various multiple finger-like structures.
These structures originate from specific boundary conditions
at $y=\lambda$. 
The existence of such complicated patterns is quite
encouraging, especially if we want to proceed further to the production,
development, competition and splitting between fingers with
the time dependent equation displayed above\footnote{Beautiful
martingales similar to the twisting and whirling of plant branches 
were also found. See remarks on symmetry below eq.(\ref{time})}.

\section{Conclusions}

The contribution of the present work consists in
the development of nonlinear differential equations for the Hele-Shaw
Saffman-Taylor fingering problem.
The approach is based upon the treatment of the
stream function as a separable potential without any free phenomenological, 
adjustable parameters.
The numerical stationary profiles, match extremely well
the experimental ones, that are, in turn, quite accurately
reproduced by the phenomenological solution of Pitts\ci{pit}.
The predicted 
dependence of the asymptotic finger half-width $\lambda$ 
on the physical parameters of the problem, is also satisfactory.

For the stationary case, the finger
nonlinear equations are eqs.(\ref{diff}), and (\ref{diffs}) 
the latter expressed in terms of the arclength,
while eq.(\ref{diff2}) is the corresponding integrodifferential equation, 
of second order, for the angle as a function of arclength. These are
nonlinear and nonlocal equations.

The time-dependent equation is of the third order in space, and 
first order in time.
Eqs.(\ref{time},\ref{times}), corresponds to $\eta$ in terms
of x and the arclength respectively, 
whereas eq.(\ref{time2}) is its integrodifferential
equation for the angle as a function of the arclength also.

The equations found here, resemble 
the Korteweg-deVries (KdV) equation, that has been
long recognized as a solitary wave equation\ci{K-dV}.
This equation was obtained for
the nonlinear propagation of shallow water waves in a channel
, and it is of the same order in space
and time, as our equations.

The KdV equation reads

\bea\label{kdv}
c_0\bigg(\frac{h^2}{6}-\frac{2~T~h}{\rho~g}\bigg)
\eta_{xxx}+\frac{3~c_0}{2~h}\eta\eta_x+\eta_x(c_0+\frac{\eta_t}{\eta_x})=0
\eea

where {\sl h}, is the depth of the channel in which the soliton propagates.
$c_0=\sqrt{g~h}$ is the speed of linear propagation of waves in the channel,
 {\sl g} is the acceleration of gravity and $\eta$ denotes the soliton
profile propagating in the channel along the {\sl x} axis with velocity 
$\ds U~=~-\frac{\eta_t}{\eta_x}$,
 in a fluid with density $\rho$ and surface tension {\sl T}.
This is a third order equation nonlinear in $\eta$. 
The specific nonlinear term, and,
the potential found in eq.(\ref{diff}), differ from the quadratic
nonlinear term of the KdV equation (\ref{kdv}).
The nonlinear terms in eq.(\ref{diff}), and eq.(\ref{kdv})
arise from the boundary conditions at the interface between the front 
and the displaced fluid as well as the surface tension pressure drop.
Eq.(\ref{diff}) however, carries also the information of the lateral
flow constraint that is absent in eq.(\ref{kdv}).
Moreover, eq.(\ref{diff}) is based upon the creeping flow assumption of eq.
(\ref{creep}), while eq.(\ref{kdv}) does not assume such
a restricted version of the Navier-Stokes equation. Therefore
the nonlinear terms differ.

Nevertheless, the equations are similar. Both are of third order
in the shape variable,  both are nonlinear, and are linear in time 
In the derivation of
eq.(\ref{diff}) we adhered to the separable assumption expansion
used by Korteweg and deVries\ci{K-dV} and their techniques. 
It is then, perhaps, not surprising, that we found a similar equation
- despite the differences in context-, as both equations 
describe the propagation of stationary and
 stable shape fronts in a fluid medium.

Links between soliton equations and the Hele-Shaw finger are already hinted
in the work of Kadanoff\ci{kad}, in which varieties of the so-called
Harry-Dym equations were found to be in close relationship to
finger development.
The above equations belong to a broad class
of {\sl flux-like} partial differential equations, that
 are widely used in the literature. In the context of Darcian
flow in the Hele-Shaw cell,
 Goldstein et al.\ci{gold} study instabilities and singularities
by means of an equation of this type.
The main difference between the equations we derived here, and the one
of Goldstein et al.\ci{gold}, is, again, the appearance of
a potential term $W(\eta)$ in eqs.(\ref{time},\ref{time2}).

The present method does not use
any phenomenologically observed characteristic. The
model results follow solely from a treatment in terms of the stream function,
and the relaxation of the condition of $\lambda=\frac{U}{V}$, that
amounts to an unphysical restriction on the Hele-Shaw cell walls
especially for large viscosities, or small capillary numbers.
A flat tail requires
$\ds \epsilon=\frac{5~\lambda^2-1}{4~\lambda^2}$. 
$\epsilon$, is then determined by $\lambda$, that is in turn
 determined by the existence of solutions. 
We assumed a constant, capillary number dependent $\eps$.
Further progress requires a model for
the dependence of $\eps$ on the capillary number.

The predicted dependence of finger half-width $\lambda$,
 on capillary number through the parameter {\sl B} of figure 4,
achieves more than the integrodifferential approach of
McLean and Saffman\ci{mc} even without corrections
of finite films left behind the propagating finger.
The partial success in dealing with the problem, 
suggests that front propagation phenomena\ci{pel} 
might have a corresponding soliton-like guiding equation, 
at least for restricted geometries.

For the Hele-Shaw cell, 
the cell provides the potential for the propagation of the
finger, while the pressure difference determined by the surface tension
provides the dynamics.
In dendritic growth, the temperature gradient 
provides the dynamical guidance and the cell and substrates, the potential.

The obvious shortcoming of the current 
method, and other existing methods, is the
neglect of the dynamics along the third dimension. 
One possible way to improve the equations would be to include vorticity.
Even if in the bulk there is
no vorticity, on the front there could be a sizable amount of it without
spoiling the bulk properties of the fluid in the hydrodynamic regime, 
as in superfluid Helium\ci{till}.
It seems then, that a treatment including vorticity may be more appropriate.

\end{document}